\pdfoutput=1
\documentclass[iop,twocolumn,jphysc,10pt,showpacs,floatfix,nofootinbib,
superscriptaddress
]{revtex4-1}
\usepackage{subfigure}
\usepackage{amssymb}
\usepackage{amsfonts}
\usepackage{amsmath}
\usepackage{amsthm}
\usepackage{epsfig}
\usepackage{graphicx}
\usepackage[usenames,dvipsnames]{color}
\usepackage[latin1]{inputenc}
\usepackage{hyperref}
\usepackage{comment}

\begin{document}

\title{Interplay between morphological and shielding effects in field emission via Schwarz-Christoffel transformation}

\author{Edgar Marcelino}
\email{edgarufba@gmail.com}
\address{Centro Brasileiro de Pesquisas F\'{i}sicas, Rua Dr. Xavier Sigaud 150, 22290-180, Rio de Janeiro, RJ, Brazil}

\author{Thiago A. de Assis}
\email{thiagoaa@ufba.br}
\address{Instituto de F\'{\i}sica, Universidade Federal da Bahia,
   Campus Universit\'{a}rio da Federa\c c\~ao,
   Rua Bar\~{a}o de Jeremoabo s/n,
40170-115, Salvador, BA, Brazil}

\author{Caio M. C. de Castilho}
\email{caio@ufba.br}
\address{Instituto de F\'{\i}sica, Universidade Federal da Bahia,
   Campus Universit\'{a}rio da Federa\c c\~ao,
   Rua Bar\~{a}o de Jeremoabo s/n,
40170-115, Salvador, BA, Brazil}
\address{Instituto Nacional de Ci\^{e}ncia e Tecnologia em Energia e Ambiente - INCTE\&A, Universidade Federal da Bahia,
   Campus Universit\'{a}rio da Federa\c c\~ao,
   Rua Bar\~{a}o de Jeremoabo s/n,
40170-280, Salvador, BA, Brazil}
\address{Centro Interdisciplinar em Energia e Ambiente, Universidade Federal da Bahia,
   Campus Universit\'{a}rio da Federa\c c\~ao, 40170-115, Salvador, BA, Brazil}

\begin{abstract}
It is well known that sufficiently strong electrostatic fields are able to change the morphology of Large Area Field Emitters (LAFEs). This phenomenon affects the electrostatic interactions between adjacent sites on a LAFE during field emission and may lead to several consequences, such as: the emitter's degradation, diffusion of absorbed particles on the emitter's surface, deflection due to electrostatic forces and mechanical stress. These consequences are undesirable for technological applications, since they may significantly affect the macroscopic current density on the LAFE. Despite the technological importance, these processes are not completely understood yet. Moreover, the electrostatic effects due to the proximity between emitters on a LAFE may compete with the morphological ones. The balance between these effects may lead to a non trivial behavior in the apex-Field Enhancement Factor (FEF). The present work intends to study the interplay between proximity and morphological effects by studying a model amenable for an analytical treatment. In order to do that, a conducting system under an external electrostatic field, with a profile limited by two mirror-reflected triangular protrusions on an infinite line, is considered. The FEF near the apex of each emitter is obtained as a function of their shape and the distance between them via a Schwarz-Christoffel transformation. Our results suggest that a tradeoff between morphological and proximity effects on a LAFE may provide an explanation for the observed reduction of the local FEF and its variation at small distances between the emitter sites. \end{abstract}

\maketitle

\section{Introduction}
The emission of electrons by a conducting surface, when a strong electrostatic field is applied, is an interesting phenomenon that has led to important scientific and technological developments \cite{Muller1937,Muller1951,Muller1956,Edgcombe,Cole2015chapter}, specially in the particular regime of Cold Field Electron Emission (CFE) \cite{Jeffreys,FowlerN,Burgess,MG,Forbes,ForbesJPhysA,Forbes2013}. Although this is a quantum phenomenon, since it is explained by a tunneling process, it also involves a classical counterpart. This happens because the potential barrier experienced by the electrons during the emission depends on the electric field on the surface, which is determined from the solution of Laplace's equation. For this reason, studying classical solutions of the electrostatic field on surfaces with different geometries is also an important task to describe the field emission phenomenon. Since very high fields are required to extract electrons, i.e. of the order of a few V/nm in pure metals with a classical planar smooth surface, geometries able to provide a large local Field Enhancement Factor (FEF) are then required. Thus, it is interesting to consider emitters with geometries having protrusions, corners, edges and/or tips, such as the ones proposed in Refs. \cite{Ryan1,Ryan2,Shiffler1,Shiffler2,Jones,deAssisJAP,Marcelino2017}. The phenomenon of higher emission near the emitters' edges is an important topic surveyed in Ref.\cite{PENGAPR2017}. The FEF at the apex of a protrusion is a particularly important physical quantity \cite{Edgcombe}.

The interest in producing Large Area Field Emitters (LAFEs) has substantially increased in order to achieve a better understanding in the production of vacuum nanoelectronic devices for applications such as: high-brightness electron sources \cite{JensenPhysRev}, high power microwave vacuum devices \cite{Spindt2000} and x-ray generators \cite{Xray}. Examples of reported field emission cathodes for x-ray sources include silicon \cite{Xray1} and carbon nanotube emitter arrays \cite{Xray1}. The sub-critical self-oscillation of a field emission nanoelectromechanical system, formed by a single resonator, has also been observed \cite{Ayari2007}. This opens new possibilities for the development of high-speed autonomous nanoresonators and signal generators, showing that field emission is a promising physical route for building new nanocomponents.

In a LAFE, the emitter comprises of many individual emitters or emission sites, each one with its own characteristics. Considering ungated emitters, two aspects are particularly relevant for a LAFE's production. First, the actual spacing between the emitters in an array will vary from their nominal spacing due to limitations in manufacturing devices/processes \cite{Jensen2017}. Second, near the border of the array, the emitters experience a reduced electrostatic interaction that leads to higher FEFs than the ones at the center of the array. This is responsible for a nonuniform emission along the LAFE \cite{Harris2016}. These aspects have motivated theoretical studies based on the electrostatic interaction between emitters in small clusters \cite{Jensen2015,JensenAPL2015,Jensen2016AIPA,RFJAP2016,ForbesAssis2017,FT2017JPCM}. In addition, an interesting phenomenon, called Close Proximity Electrostatic Effect (CPEE), has been recently reported \cite{Jensen2015,JensenAPL2015,ForbesAssis2017,FT2017JPCM}. It is generally expected that the FEF of two emitter sites, in a small cluster or in an array, would be reduced when they are close to each other. The CPEE, differently from the electrostatic shielding, consists of an increasing of the characteristic local FEF, as long as the emitter sites become closer to each other in some specific interval \cite{Jensen2015,JensenAPL2015,ForbesAssis2017,FT2017JPCM}. Since the CPEE was recently reported and there are just a few works concerning this effect in the literature, it is not yet clear whether any kind of geometry would be able to provide CPEE. Thus, this investigation in itself is appealing, considering that this effect is promising for technological applications, specially the ones related to self-oscillations of mult-tip resonators.

Another difficulty to be overcome during LAFE's applications, which includes electron sources, is the vacuum arcing. A common hypothesis behind this phenomenon, that usually occurs after intense field electron emission, is the emergence of high local current densities, yielding the so called Nottingham heating \cite{Nottingham0,Nottingham}. Nottingham heating leads to local temperatures close to the melting point of the emitter and evaporation of neutral atoms \cite{Dyke,Dyke2,Kyritsakis}. As a consequence of this phenomenon, the  shape of the emitter sites  and the distance between them may vary, leading to a possible degradation of the emitter. These aspects are expected to modify the corresponding local FEFs along the LAFE, by a competition between two effects: the changing in the morphology of the emitter sites and the reduction of their local FEFs due to the proximity between them, the last one usually known by ``shielding". Cahay and collaborators have reported that the Fowler-Nordheim (FN) plots reflect the changing in the FEF when the emitter's morphology, formed by carbon nanotube (CNT) fibers, varies when a high field is applied \cite{Cahay2014}. A few years later, the influence of the morphology of a single CNT in field emission has been studied for different values of the cathode-anode gap \cite{PengAIP2017}. The influence of vacuum breakdown on the behavior of the emitter's morphology has been considered in \cite{ieee2014}. Also, the electric field distribution and current emission in a miniaturized geometrical diode has been recently studied \cite{JAP2017SC}.

Following the aforementioned motivations, this paper provides an analytical study of the apex-FEF of two identical emitters mirror-symmetrically disposed, that consider the interplay between shape and shielding effects during field emission. To this end, we have considered a two-dimensional emitter profile, with its frontier limited by two identical triangular protrusions on a line, under an external electrostatic field. The FEF in the vicinity of the top of each triangular protrusion is analytically obtained, as a function of the geometric parameters of the LAFE, via Schwarz-Christoffel conformal mapping. The case of two adjacent triangular protrusions is considered at first and then the results are generalized to the case of distant protrusions. The results presented in this manuscript allow one to infer the limits in which the increase of the FEF, due to a variation of the emitters' shape, is attenuated by shielding, as the distance between the protrusions decreases. Finally, it is showed that, when the triangular protrusions considered here are sufficiently close to each other, a fast decrease of the FEF and its non trivial variation is observed. This is expected to yield a significant reduction in the local current density and its variation, yielding the so called saturation of the corresponding FN plots. Our results show that there is a competition between shielding and morphological effects, which may yield a FEF behavior similar to the one obtained for the CPEE, although this effect does not exist here. One should notice that, despite the similarities to the CPEE, applications of our results to self-oscillations and multi-tip resonators are not trivial to be done. One reason for this is that the conformal mapping technique used here is effective only for two-dimensional geometries, such as the ridge emitters considered here. This kind of geometry differs considerably, for instance, from the typical ones of nanotubes used in resonator devices. For this reason, the present work focuses only on the degradation and tradeoff between proximity and morphological effects in a LAFE.

The sequence of the paper is organized as follows. In
Sec. II, the analytical expression for the apex-FEF, considering a single scalene triangular protrusion, is derived. In Sec. III, the apex-FEF of
two adjacent triangular protrusions is obtained and the consequences of changing the field emitter's shape are studied. Section IV generalizes the results of Sec. III to the case of distant triangular protrusions and presents the discussions considering the tradeoff between  variation of the emitter's shape and shielding under close proximity.
Section V summarizes the conclusions of this work.

\section{FEF in the case of a single triangular protrusion}

\begin{figure}[h!]
\includegraphics [width=6.0cm,height=5.0cm] {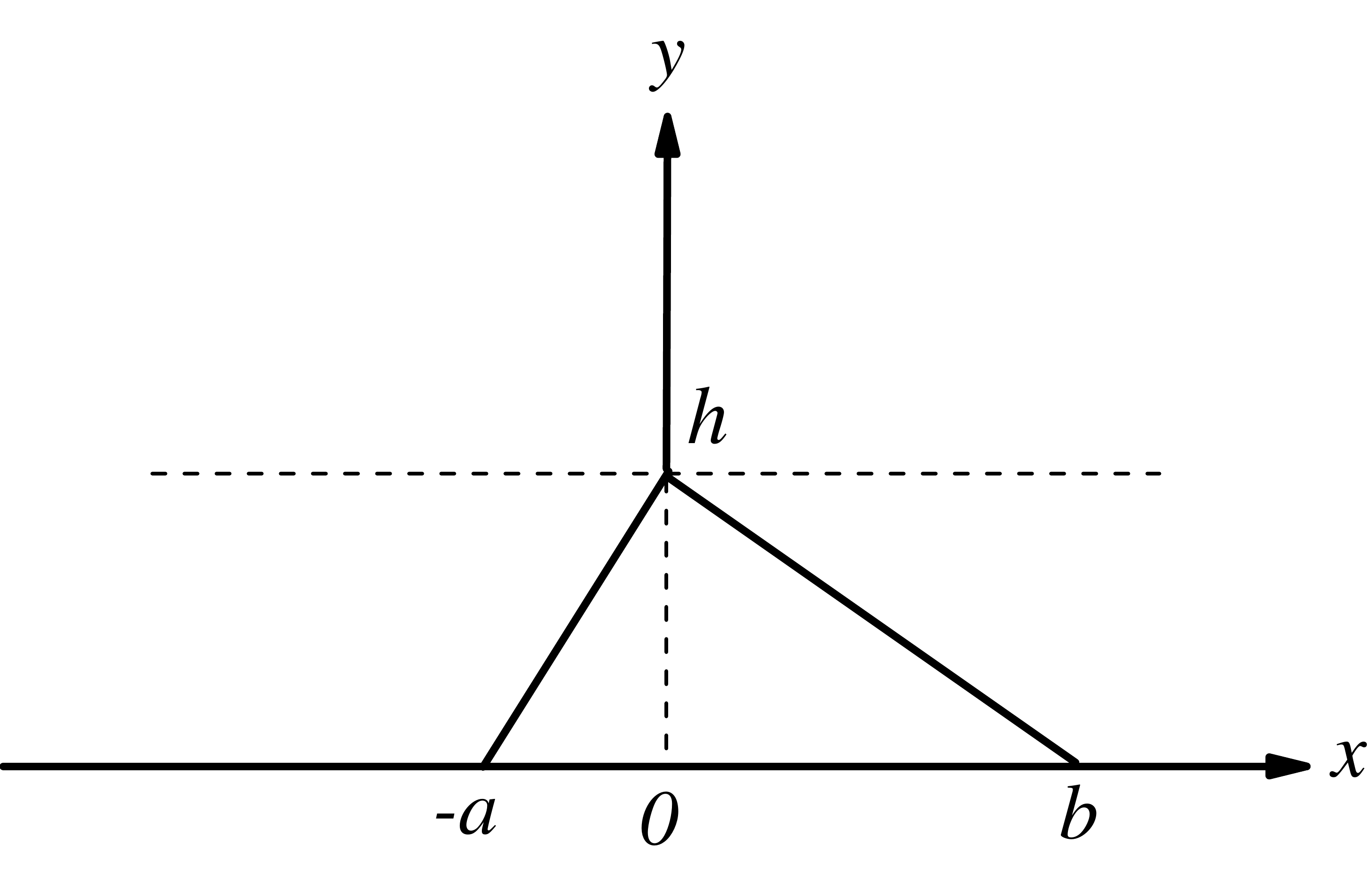}
\caption{Single scalene triangular protrusion on a conducting line: z-plane.} \label{ST}
\end{figure}

Before considering the case of two protrusions, we derive the solution for the apex-FEF of a conducting profile limited by a  single triangular protrusion on an infinite line under an external electric field. In summary, the solution obtained in \cite{Marcelino2017} for an isosceles triangle is now generalized to the case of a scalene triangular protrusion in this section, as shown in Fig. \ref{ST}.

The Schwarz-Christoffel transformation [\onlinecite{Churchill},\onlinecite{Hilderbrand}] mapping the u-axis in the $w=(u,v) \rightarrow u+iv$ plane into the polygonal line in Fig. \ref{ST} in the $z=(x,y) \rightarrow x+iy$ plane is given by:
\begin{equation}  \label{SCmap}
z(w)=A \int_{z_0}^{w} \frac{w^{\alpha+\beta} dw}{(w+1)^{\alpha}(w-1)^{\beta}}+C,
\end{equation}
where $\alpha=\frac{1}{\pi} arctan \left( \frac{h}{a} \right)$, $\beta=\frac{1}{\pi} arctan \left( \frac{h}{b} \right)$ and the points $(-1,0)$, $(1,0)$ and $(0,0)$ in the $w=(u,v) \rightarrow u+iv$ plane are respectively mapped into the following points in the $z=(x,y) \rightarrow x+iy$ plane: $(-a,0)$, $(b,0)$ and $(0,h)$. After choosing $z_{0}=0$, these map correlations allow one to obtain:
\begin{equation}
C=ih
\end{equation}
and
\begin{multline}
A=\frac{\left[B(\alpha+\beta+1,1-\alpha) \right]^{-1} \sqrt{h^{2}+a^{2}}}{  _{2}F_{1}(\beta,\alpha+\beta+1,\beta+2,-1)}= \\
=\frac{\left[ B(\alpha+\beta+1,1-\beta) \right]^{-1} \sqrt{h^{2}+b^{2}}}{ _{2}F_{1}(\alpha,\alpha+\beta+1,\alpha+2,-1)}.
\end{multline}
In the previous equation, the beta function $(B)$ and the following integral representation of the hypergeometric function $(_{2}F_{1})$ were used:
\begin{equation}
B(b,c-b) _{2}F_{1}(a,b,c,z)= \int_{0}^{1} \frac{x^{b-1} (1-x)^{c-b-1} }{(1-zx)^{a}} dx.
\end{equation}

Considering the $w$-plane, the complex electric potential, that remains constant along the u-axis and yields a uniform field $\mathbf{E_0}$ far away from the emitter, is given by $\phi(w)=iA |\mathbf{E_{0}}|w$. Thus, the electric field in the z-plane obeys the following expression:
\begin{equation} \label{EFieldST}
|E_{x}-iE_{y}|=\left| \frac{d \phi /dw}{dz/dw} \right|= \frac{ |\mathbf{E_{0}}| |w+1|^{\alpha}|w-1|^{\beta} }{|w|^{\alpha+\beta}}.
\end{equation}
In the vicinity of the point $z=ih$ $(w \approx 0)$, Eq. (\ref{SCmap}), with the values of $A$ and $C$ already found, can be approximated to:
\begin{equation}
|z(w)-ih|\approx \frac{\sqrt{h^{2}+b^{2}}}{ \xi(\alpha,\beta)} \frac{|w|^{\alpha+\beta+1}}{\alpha+\beta+1},
\end{equation}
where $\xi(\alpha,\beta)=\frac{ _{2}F_{1}(\alpha,\alpha+\beta+1,\alpha+2,-1)}{[B(\alpha+\beta+1,1-\beta)]^{-1}}$. Eq. (\ref{EFieldST}) can also be similarly simplified:
\begin{equation}
|E_{x}-iE_{y}|=|\mathbf{E_{0}}| \left[ \frac{\sqrt{h^{2}+b^{2}} |z-ih|^{-1}}{(\alpha+\beta+1)\xi(\alpha,\beta)} \right].
\end{equation}
By using the last equations, one can finally obtain the expression for the FEF $\left( \gamma(x,y) \equiv \frac{|\mathbf{E}|}{E_0} \right)$ near the upper corner of the triangular protrusion:
\begin{equation} \label{FEFST}
 \gamma(x,y) \approx \left[\frac{\sqrt{h^{2}+b^{2}}}{(\alpha+\beta+1)\xi(\alpha,\beta) \Delta(x,y)} \right]^{\frac{\alpha+\beta}{\alpha+\beta+1}},
\end{equation}
where $\Delta(x,y)=\sqrt{x^{2}+(y-h)^{2}}$ is the distance from the top of the emitter to the point $(x,y)$, where the FEF is evaluated. This expression is used to plot the dashed lines in Figs. \ref{AdD}, \ref{DistD} and \ref{Distdd}, which will be discussed in Sec. III and Sec. IV.

Equation (\ref{FEFST}) becomes more accurate as long as the distance to the singularity (apex corner) decreases, since it is a consequence of a first-order approximation in a series expansion near the corner $(w=0)$. Thus, this analytical approach provides an effective method to derive the local FEF in the vicinity of corners and tips, differently from the numerical ones that usually fail near singularities. For large distances to the corner, the aforementioned approximation fails and higher-order terms from the series expansion become necessary. In this case, the $(x,y)$ dependence of the local FEF will not simply be a function of the distance $(\Delta(x,y))$ to the corner. In the region of validity of Eq. (\ref{FEFST}) one can consider the direction of the electric field approximately perpendicular to the emitter, in agreement with the Neumann conditions of the problem. Furthermore, the discussion in this paragraph applies to other local FEFs, derived in same way in the vicinity of corners, as the ones in Eq. (\ref{FEF1}), Eq. (\ref{FEF3}) and some of the local FEFs derived in other works in the literature \cite{Ryan1,Ryan2,Tang2011,Marcelino2017}.
\section{FEF in the case of two adjacent triangular protrusions}

\begin{figure}[h!]
\includegraphics [width=8.0cm,height=5.2cm] {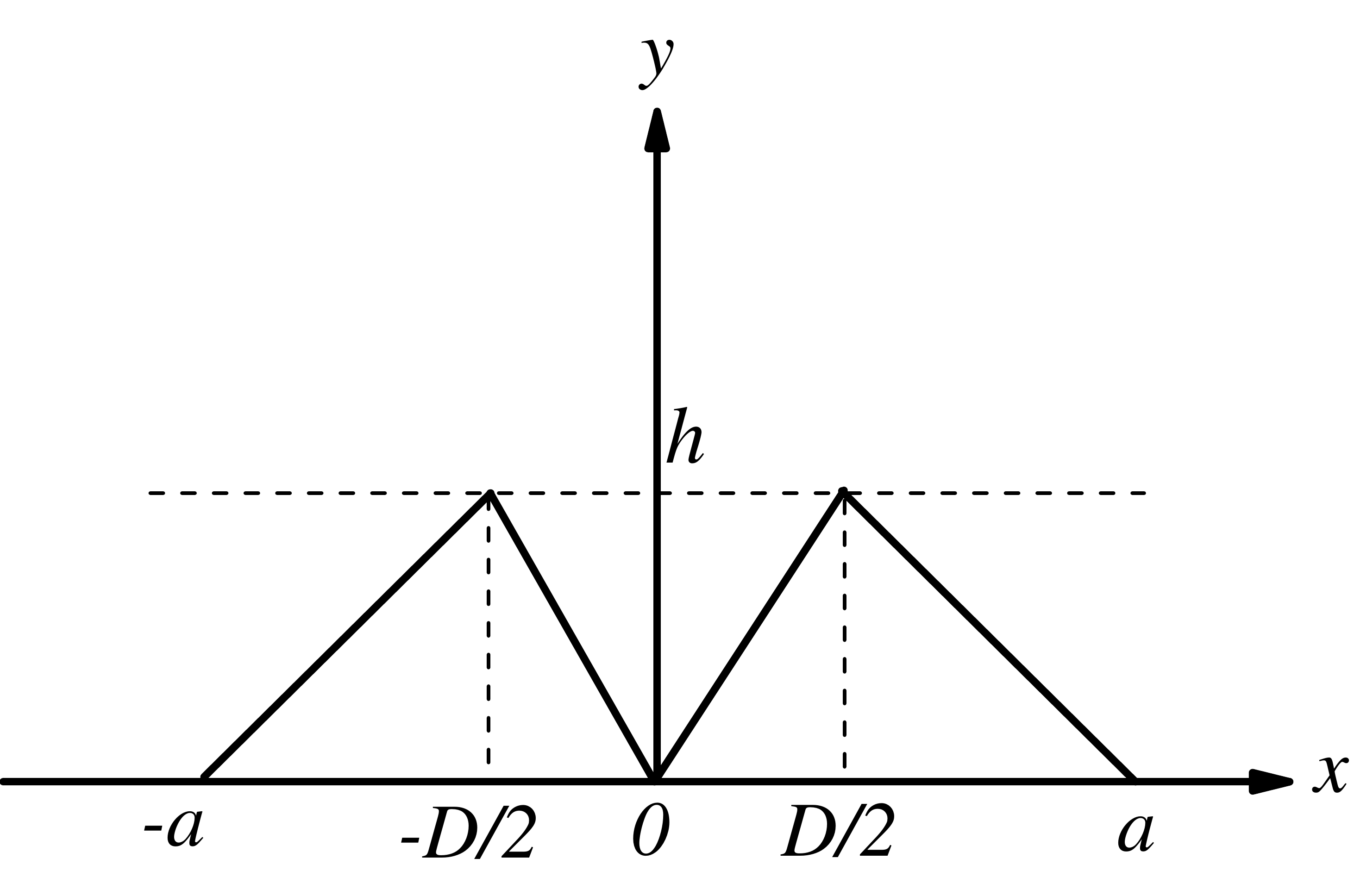}
\caption{Two adjacent mirror-reflected triangular protrusions on a conducting line: z-plane.} \label{AdT}
\end{figure}

In this section, we consider the case of a conducting profile limited by two adjacent triangular protrusions on an infinite line under an external electrostatic field, as showed in Fig. \ref{AdT} on a $z=(x,y) \rightarrow x+iy$ plane. The two triangular protrusions have a common vertex, width $a$, height $h$ and the distance between their upper vertices is $D$. To study this system, let us consider a Schwarz-Christoffel transformation mapping the $u$-line and the region above in the $w=(u,v) \rightarrow u+iv$ plane into the polygonal line with the emitter's shape and the region above in the $z=(x,y) \rightarrow x+iy$ plane, showed in Fig. \ref{AdT}. This transformation is given by the following formula:
\begin{equation} \label{SC1}
z(w)=A \int_{0}^{w} \frac{(w^{2}-1)^{\alpha+\beta} dw}{(w^{2}-u^{2})^{\beta} w^{2 \alpha}} +C,
\end{equation}
where the exponents $\alpha$ and $\beta$ are determined from the angles of the polygonal line in the z-plane, $\alpha= \frac{1}{\pi} \arctan \left( \frac{2h}{D} \right)$ and $\beta= \frac{1}{\pi} \arctan \left( \frac{2h}{2a-D} \right)$. This transformation fulfills the following correlations between specific points in the w- and z-planes: $w=(0,0) \leftrightarrow z=(0,0)$, $w=(\pm 1,0) \leftrightarrow z=(\pm \frac{D}{2},h)$ and $w=(\pm u,0) \leftrightarrow z=(\pm a,0)$. The first two of these correlations yield:
\begin{equation}
C=0
\end{equation}
and
\begin{eqnarray}
A=\frac{u^{2 \beta} \left[ B \left(\frac{1}{2}-\alpha,\alpha+\beta+1 \right) \right]^{-1} \sqrt{D^{2}+4h^{2}}}{  _{2}F_{1} \left(\beta,\frac{1}{2}-\alpha,\beta+\frac{3}{2},u^{-2} \right)}. \label{FindA}
\end{eqnarray}
The third correlation, combined with the last equation, specifies $u$ as a solution of the following equation:
\begin{equation}
\frac{2 u^{2 \beta} \int_{1}^{u} \frac{(w^{2}-1)^{\alpha+\beta} dw}{(u^{2}-w^{2})^{\beta} w^{2 \alpha}}}{\xi(\alpha,\beta,u)}=\sqrt{\frac{\left(\frac{a}{h}-\frac{D}{2h} \right)^{2}+1}{\left(\frac{D}{2h} \right)^{2}+1}},
\end{equation}
where $\xi(\alpha,\beta,u)=\frac{_{2}F_{1} \left( \beta,\frac{1}{2}-\alpha,\beta+\frac{3}{2},u^{-2} \right)}{\left[ B \left(\frac{1}{2}-\alpha,\alpha+\beta+1 \right)  \right]^{-1}}$.


%

According to the discussion in Sec. II, the electric field in the z-plane can be expressed by:
\begin{equation} \label{AdEF}
E_{x}-iE_{y}=\frac{d \phi}{dz}=\frac{d \phi /dw}{dz/dw}=\frac{i |\mathbf{E_{0}}| (w^{2}-u^{2})^{\beta} w^{2 \alpha}}{(w^{2}-1)^{\alpha+\beta} }.
\end{equation}
In the vicinity of the top of the right triangular protrusion in Fig. \ref{AdT} $(w\approx 1)$, the following approximation is valid: $\frac{(w^{2}-1)^{\alpha+\beta}}{(w^{2}-u^{2})^{\beta} w^{2 \alpha}} \approx \frac{2(w-1)^{\alpha+\beta}}{(1-u^{2})^{\beta}}$. Thus, Eq. (\ref{SC1}) and Eq. (\ref{AdEF}), with $C=0$, reduce to the following equations:
\begin{eqnarray}
\left| z-\left( \frac{D}{2}+ih \right) \right|=\frac{2^{\alpha+\beta}A |w-1|^{\alpha+\beta|1}}{(\alpha+\beta+1)|u^{2}-1|^{\beta}}, \label{Aux1} \\
\gamma(w \approx 1) \equiv \frac{|\mathbf{E}|}{E_0}=\frac{|u^{2}-1|^{\beta}}{2^{\alpha+\beta} |w-1|^{\alpha+\beta}}.  \label{Aux2}
\end{eqnarray}

\begin{figure}[h!]
	\includegraphics [width=7.8cm,height=5.8cm]{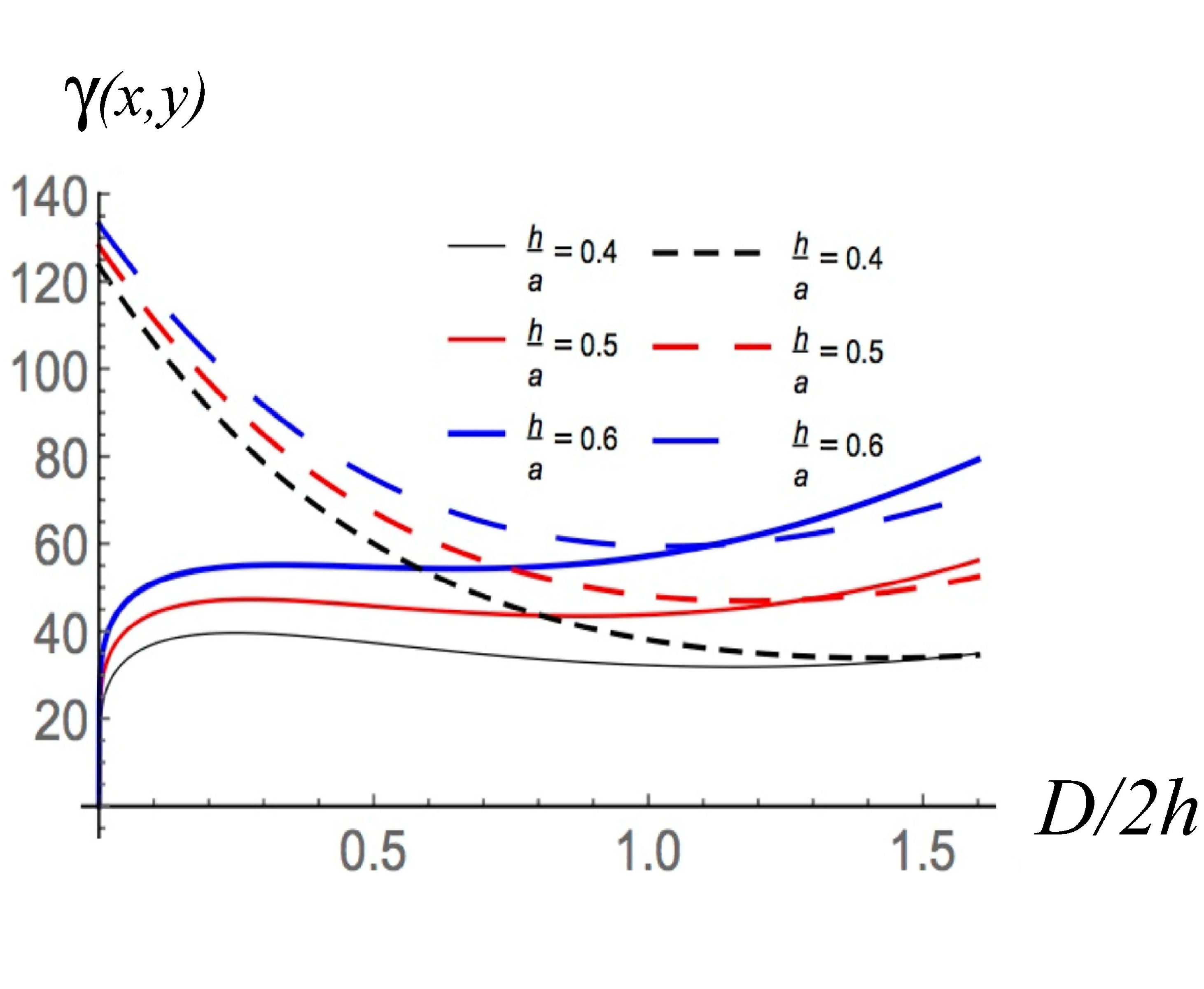}
	\caption{(Color online) The solid lines refer to the FEF near the top of the right triangular protrusion in Fig. \ref{AdT} as a function of the ratio $D/2h$ for different values of the aspect-ratio $h/a$ (ratio height-to-width) and $\Delta(x,y)/h=10^{-5}$. The dashed ones refer to the FEF near the appex of a single triangular protrusion on a line with the same shape.}
	\label{AdD}
\end{figure}

Combining Eq. (\ref{FindA}), Eq. (\ref{Aux1}) and Eq. (\ref{Aux2}), one can derive the expression for the FEF near the apex of the right triangular protrusion:
\begin{equation} \label{FEF1}
\gamma(x,y) \approx \left[ \frac{2 u^{2 \beta} \sqrt{\left(\frac{D}{2h} \right)^{2}+1}}{ (\alpha+\beta+1)  \xi(\alpha,\beta,u) \frac{\Delta(x,y)}{h}} \right]^{\frac{\alpha+\beta}{\alpha+\beta+1}},
\end{equation}
where $\Delta(x,y)=\sqrt{\left(x-\frac{D}{2} \right)^{2}+(y-h)^{2}}$ is the distance to the top vertex of the right triangular protrusion. In Fig. \ref{AdD}, the FEF near the top vertex of the protrusion (solid lines) is plotted as a function of the ratio $D/2h$ for different values of the aspect-ratio $h/a$ (ratio height-to-width). The FEF presents a local maximum as a function of the distance between the two upper vertices of the protrusion. Thus, there is a region in which the FEF increases when the distance between the peaks decreases. The dashed lines show the FEF near the top of a single triangular protrusion with the same shape on a line (with projections of the protrusion on the line with dimensions $D/2$ and $a-D/2$), in this case the FEF presents a local minimum instead of a maximum.

For small distances between the apexes $(D)$, the results presented here feature a local maximum in the apex-FEF as a function of $(D)$. This same kind of behavior was obtained for the CPEE curves, although in the CPEE the shape of the emitter does not vary \cite{Jensen2015,JensenAPL2015,ForbesAssis2017,FT2017JPCM}. Thus, the CPEE does not occur for the results derived in this section,  despite the resemblance between the shape of the curves obtained in both cases. The plot behavior presented in this section is simply a consequence of the interplay between the shielding and morphological effects in the FEF, as long as the distance between the apexes vary. Nevertheless, one should keep in mind that, although the distance between the two peaks vary, the emitters are still adjacent. In Sec. IV we discuss the possibility of CPEE in the case where the distance between the emitters also varies.

For higher values of the aspect-ratio (ratio height-to-width) of the protrusions, the effect of the shape is attenuated and the monotonic screening of the electric field tends to be recovered. The same kind of screening is obtained in \cite{Tang2011}. Interestingly, for $D/2h < 0.1$, the FEF near the apexes of the emitters decreases faster when they become close to each other. This is an interesting phenomena that can justify the saturation in the corresponding FN plots in LAFEs, since shielding is the dominant effect at small distances.  Thus, this phenomenon can constitute the origin of a sudden failure to emit for some of the emitters forming a LAFE. When the emitter sites in a LAFE are too distant from each other, the voltage-dependent FEF reduction and the voltage-divider effect, due to measurement-circuit resistance, may be an alternative explanation for the emission failure. This happens as a result of the fact that the electrostatic interactions between a single emitter site and the adjacent ones are negligible for large distances between them.
These limits are considered, for instance, in Ref. \cite{RFJAP2016}, in which a cubic power-law dependence of the FEF with the inverse of the distance between the sites is obtained. Notwithstanding, one should notice that this is not exactly what happens in this section because the sites are not distant from each other.

\section{FEF in the case of two distant triangular protrusions}

\begin{figure}
\includegraphics [width=8.0cm,height=5.2cm] {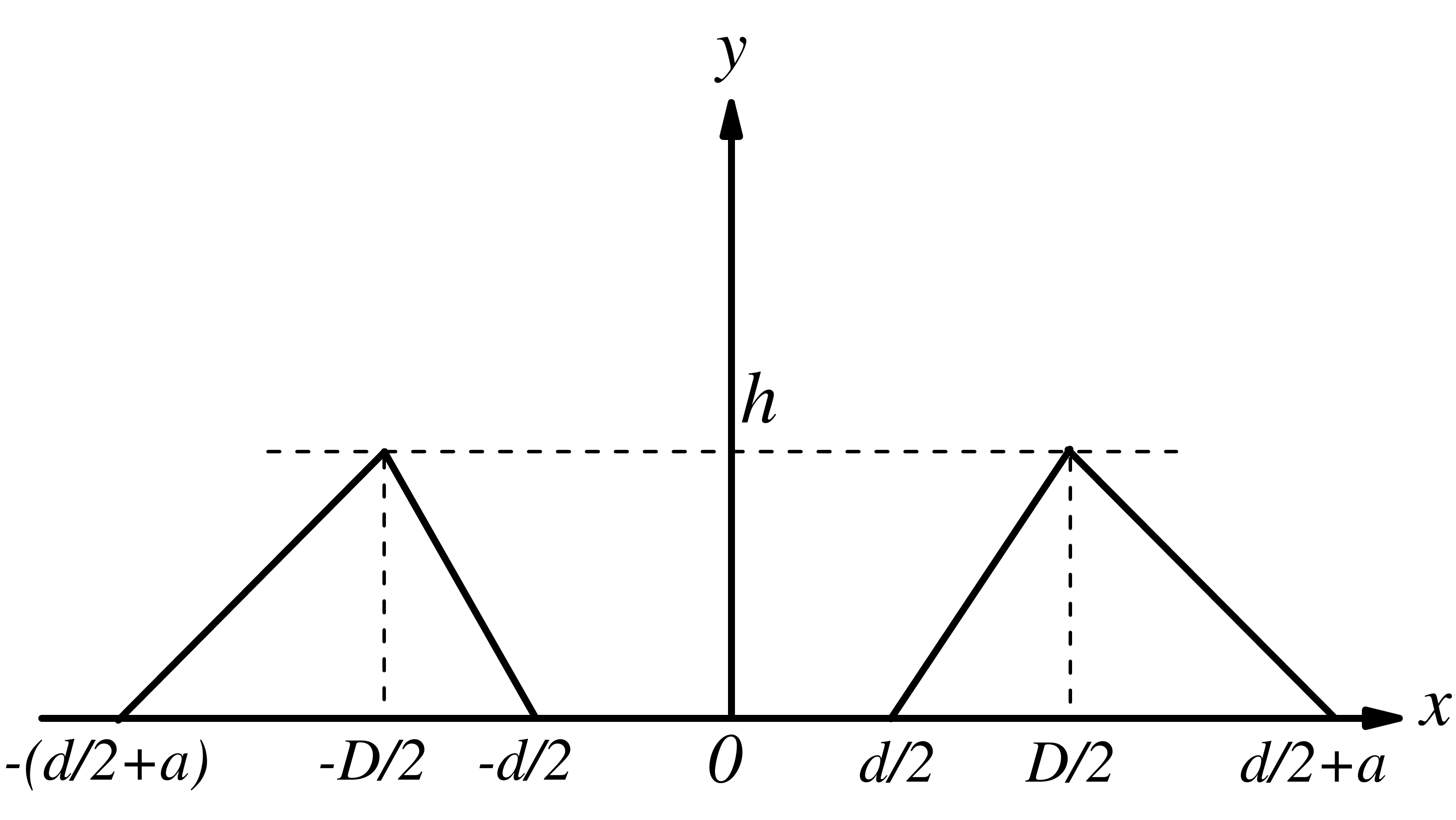}
\caption{Two distant mirror-reflected triangular protrusions on a conducting line: z-plane.} \label{DistT}
\end{figure}

Finally, we generalize the results from Sec. III to the case of a conducting profile limited by two distant triangular protrusions on a line, as shown in Fig. \ref{DistT}, when this system is under an external electric field. In this case, the Schwarz-Christoffel transformation mapping the u-axis in the w-plane into the polygonal line in Fig. \ref{DistT} is given by:
\begin{equation} \label{SC2}
z(w)=A \int_{0}^{w} \frac{(w^{2}-u^{2})^{\alpha+\beta} dw}{(w^{2}-1)^{\alpha} (w^{2}-v^{2})^{\beta} }+C.
\end{equation}
The last equation fulfills the following correlations between points in w- and z-planes:  $w=(0,0) \leftrightarrow z=(0,0)$, $w=(\pm 1,0) \leftrightarrow z=(\pm \frac{d}{2},0)$, $w=(\pm u,0) \leftrightarrow z=(\pm \frac{D}{2},h)$ and $w=(\pm v,0) \leftrightarrow z=\left(\pm \left(\frac{d}{2}+a\right),0\right)$. These correlations enable one to determine the parameters $A$, $C$, $u$ and $v$ in Eq. (\ref{SC2}). Thus, one obtains:
\begin{equation} \label{AB}
C=0 \qquad \text{and} \qquad  A=\frac{d}{2 I_{1}(u,v)},
\end{equation}
where $u$ and $v$ are solutions of the following system of equations:
\begin{equation}
\begin{cases}
I_{1}(u,v) \sqrt{\left(\frac{D}{2h}-\frac{d}{2h} \right)^{2}+1}=\frac{d}{2h} I_{2}(u,v), \\
I_{1}(u,v) \sqrt{\left(\frac{d}{2h}-\frac{D}{2h}+\frac{a}{h} \right)^{2}+1}=\frac{d}{2h}I_{3}(u,v).
\end{cases}
\end{equation}
The functions $I_{j}(u,v)$ $(j \in \{1,2,3\})$ are defined as follows:
\begin{eqnarray}
I_{1}(u,v)=\int_{0}^{1} \frac{(u^{2}-w^{2})^{\alpha+\beta} dw}{(1-w^{2})^{\alpha} (v^{2}-w^{2})^{\beta} },  \\
I_{2}(u,v)=\int_{1}^{u} \frac{(u^{2}-w^{2})^{\alpha+\beta} dw}{(w^{2}-1)^{\alpha} (v^{2}-w^{2})^{\beta} },  \\
I_{3}(u,v)=\int_{u}^{v} \frac{(w^{2}-u^{2})^{\alpha+\beta} dw}{(w^{2}-1)^{\alpha} (v^{2}-w^{2})^{\beta} }.
\end{eqnarray}
As it happens in Sec. II and Sec. III, only three points in the w-plane could be chosen to be mapped into specific points in the z-plane. The other points must be determined, after solving the system of equations for $u$ and $v$. This is an intrinsic characteristic of the Schwarz-Christoffel transformation, that can be understood as a consequence of a much broader result: the Riemann Mapping Theorem \cite{Riemann}.

\begin{figure}[h]
	\includegraphics [width=7.8cm,height=5.8cm]{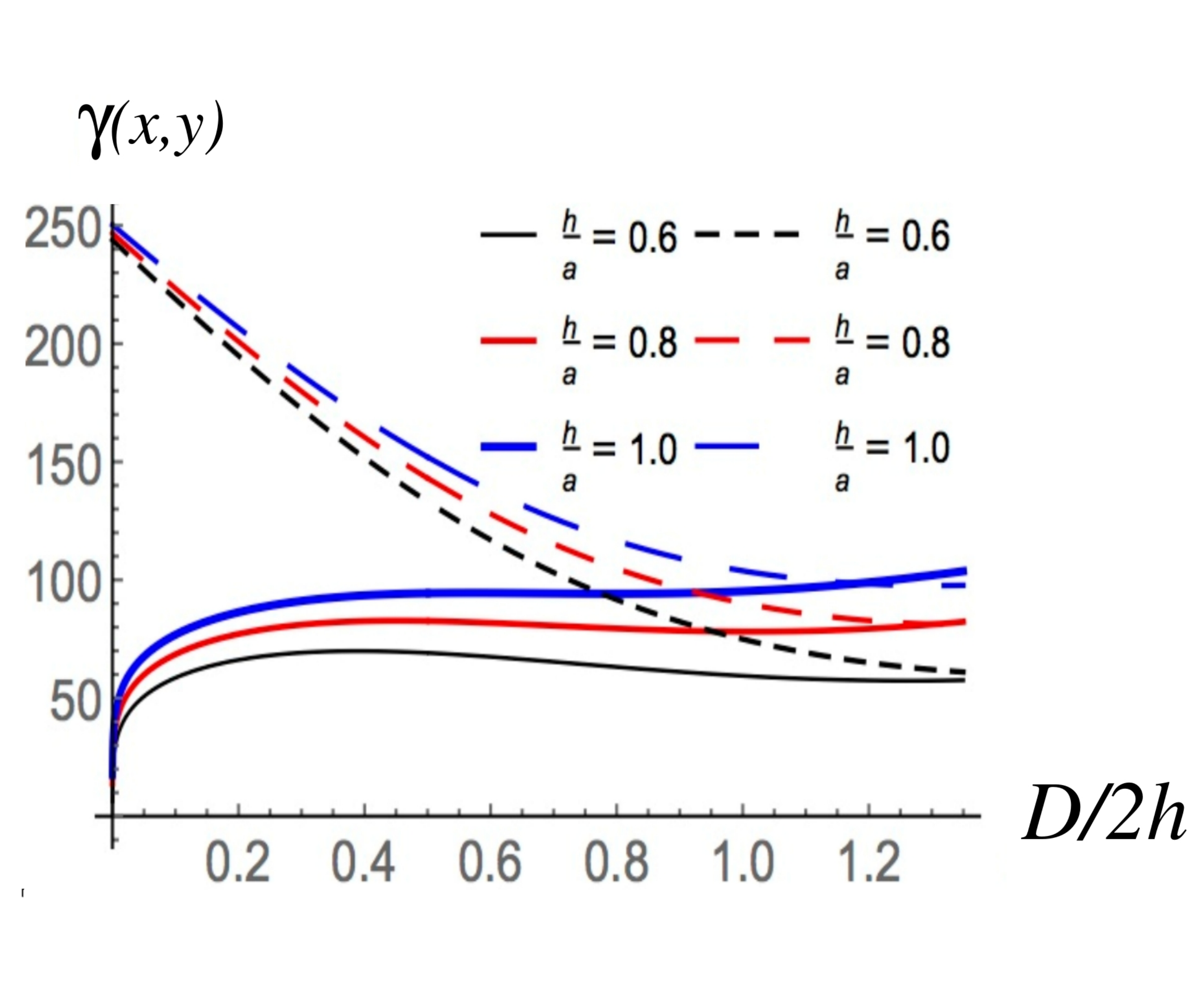}
	\caption{(Color online) FEF near the top of the right triangular protrusion, as shown in Fig. \ref{DistT}, as a function of the ratio $D/2h$ (solid lines), for different values of the aspect-ratio $h/a$ (ratio height-to-width), $d/2h=0.5$ and $\Delta(x,y)/h=10^{-5}$. The dashed lines refer to the FEF near the apex of a single triangular protrusion on a line with the same shape.}
	\label{DistD}
\end{figure}

\begin{figure}[h]
	\includegraphics [width=7.8cm,height=5.8cm]{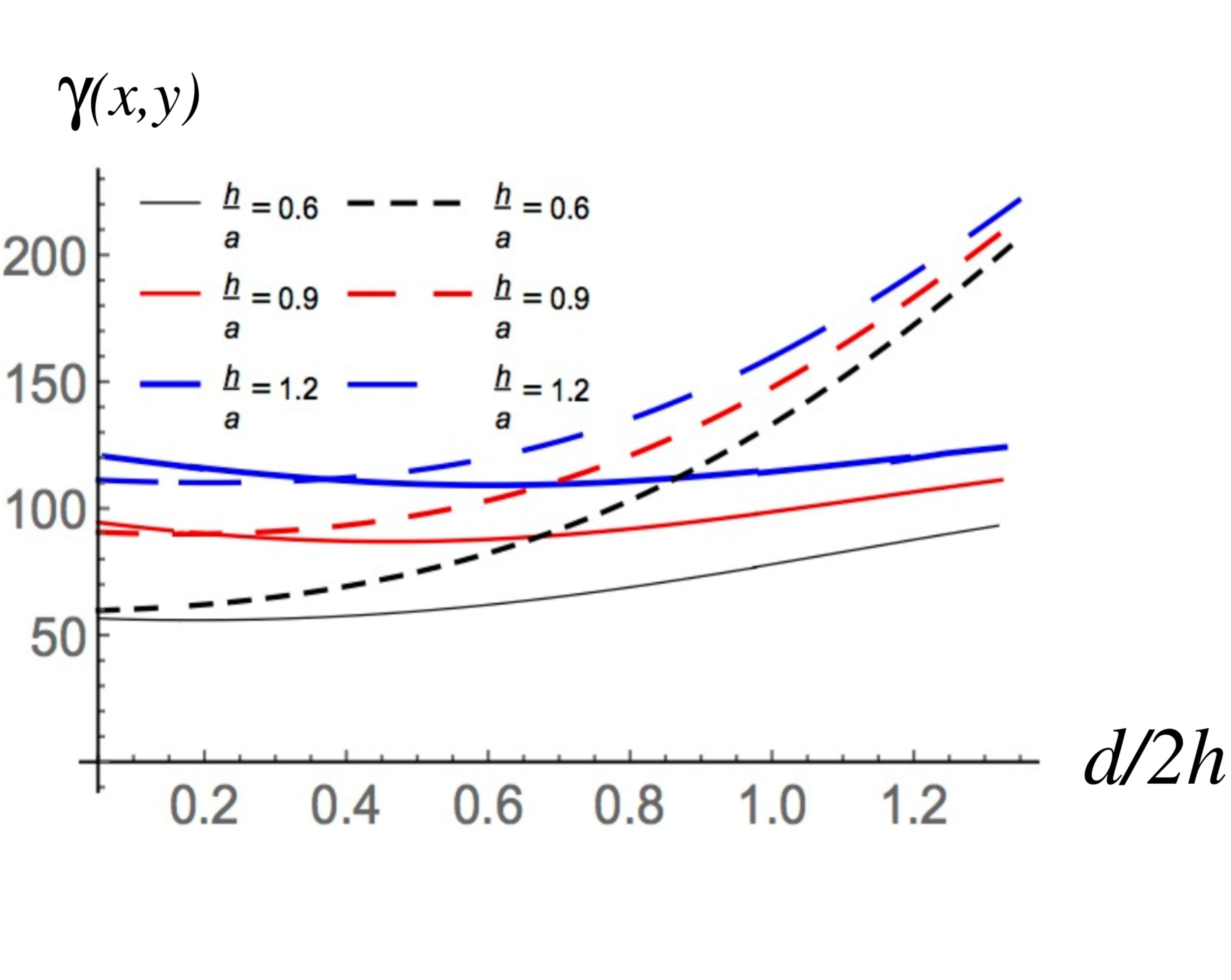}
	\caption{(Color online) FEF near the top of the right triangular protrusion, as shown in Fig. \ref{DistT}, as a function of the ratio $d/2h$ (solid lines), for different values of the aspect-ratio $h/a$ (ratio height-to-width), $D/2h=1$ and $\Delta(x,y)/h=10^{-5}$. The dashed lines refer to the FEF near the apex of a single triangular protrusion on a line with the same shape.}
	\label{Distdd}
\end{figure}

Near the top of the right triangular protrusion $(w \approx u)$, one may use the approximation given by:
\begin{equation}  \label{ApproxDist}
\frac{(w^{2}-u^{2})^{\alpha+\beta} }{(w^{2}-1)^{\alpha} (w^{2}-v^{2})^{\beta} } \approx  \frac{(2u)^{\alpha+\beta} (w-u)^{\alpha+\beta}}{(u^{2}-1)^{\alpha} (u^{2}-v^{2})^{\beta} }.
\end{equation}
Using Eq. (\ref{ApproxDist}) and following the same steps from the other sections, one may obtain:
\begin{eqnarray}
\frac{|z- \left( \frac{D}{2} +ih \right)|}{(\alpha+\beta+1)^{-1}} \approx \frac{A (2u)^{\alpha+\beta} |w-u|^{\alpha+\beta+1}}{|u^{2}-1|^{\alpha} |v^{2}-u^{2}|^{\beta}},  \label{Appr1} \\
\gamma(w \approx u) \equiv \frac{|\mathbf{E}|}{E_0}=\frac{|u^{2}-1|^{\alpha} |u^{2}-v^{2}|^{\beta} }{2u|w-u|^{\alpha+\beta}}. \label{Appr2}
\end{eqnarray}
Combining Eq. (\ref{AB}), Eq. (\ref{Appr1}) and Eq. (\ref{Appr2}), it is straightforward to obtain the FEF near the upper vertex of the right triangular protrusion $(w \approx u)$ in Fig. \ref{DistT}:
\begin{equation} \label{FEF3}
\gamma(x,y) \approx \left[\frac{\frac{d}{h} (u^{2}-1)^{\frac{\alpha}{\alpha+\beta}} (v^{2}-u^{2})^{\frac{\beta}{\alpha+\beta}}}{4u(\alpha+\beta+1) I_{1}(u,v) \frac{\Delta(x,y)}{h}} \right]^{\frac{\alpha+\beta}{\alpha+\beta+1}},
\end{equation}
where now $\Delta(x,y)=\sqrt{\left(x-\frac{D}{2}\right)+(y-h)^{2}}$ is the distance to the vertex $(D/2,h)$ in Fig. \ref{DistT}. In Figs. \ref{DistD} and \ref{Distdd}, the FEF near the apex of the right triangular protrusion (solid lines) is plotted as a function of the ratios $D/2h$ and $d/2h$ respectively.
\begin{figure}[h]
	\includegraphics [width=7.8cm,height=5.8cm]{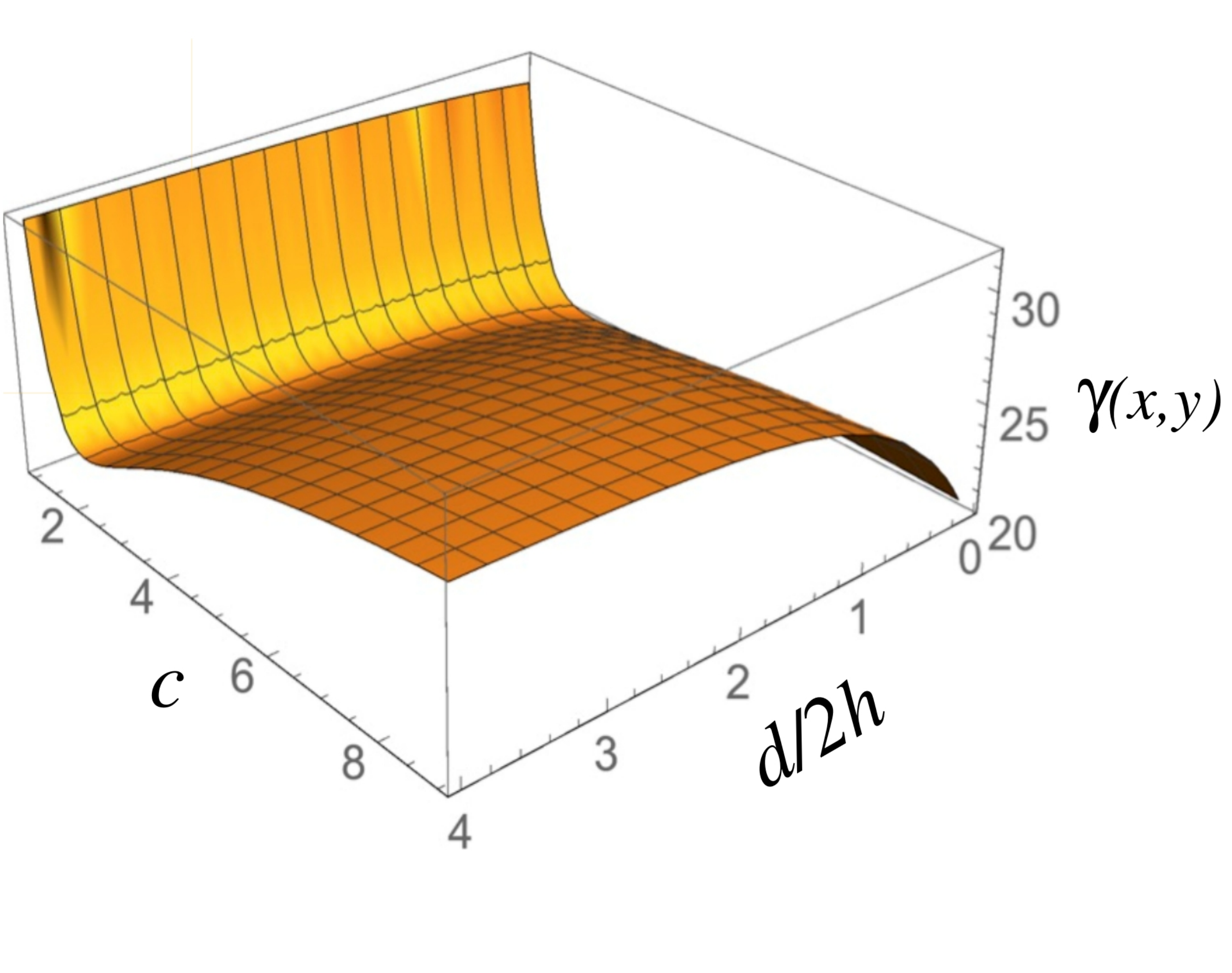}
	\caption{(Color online) FEF near the top of the right triangular protrusion in Fig. \ref{DistT} when the shape of the protrusions does not vary $(D=d+2a/c)$ as a function of the ratio $d/2h$ and the parameter $c$ for $\Delta(x,y)/h=10^{-5}$ and $h/a=0.5$. }
	\label{Dist3d}
\end{figure}

As in the previous case, a local maximum is obtained for the FEF as a function of the distance between the top of the two emitters, see Fig. \ref{DistD}. These effects are more pronounced for small values of the  aspect-ratio (ratio height-to-width). On the other hand, the behavior of the FEF with the distance between the two emitters is different, see Fig. \ref{Distdd}. In this case, a global minimum is present, which disappears for small enough values of the aspect-ratio. The FEF of a single triangular protrusion with the same shape is also plotted  in Fig. \ref{Distdd} (dashed lines) revealing that the increasing of the FEF for small distances is an effect of the change in the shape of the emitters. In Fig. \ref{Dist3d}, this FEF is plotted for the case in which the shape of the emitters does not change $(D=d+\frac{2a}{c})$ as a function of the parameter $c$ and the distance of the protrusions. In this case, the FEF increases monotonically with $d/2h$, reinforcing our conclusions.

This implies that, although the shape of the curves resembles the one from the CPEE curves in \cite{Jensen2015,JensenAPL2015,ForbesAssis2017,FT2017JPCM}, the geometry considered here is not able to provide the CPEE. It provides, in fact, a combination of the screening due to proximity presented in \cite{Tang2011}, together with the effect of the change in the shape of the emitter, as showed in \cite{Marcelino2017} for the case of an isosceles triangular protrusion on a line and in Eq. (\ref{FEFST}) for a scalene triangular protrusion.

As one varies the parameters describing the geometry of the emitter, a redistribution of the charge density occurs along the system, in order to achieve a new configuration of electrostatic equilibrium. This may happen in such a way that charge carriers may migrate from the emitter to the substrate, for equal emitters regularly spaced in the LAFE and in the system of only two emitters presented here; or from one emitter to the others, for systems consisting of small clusters of emitters \cite{ForbesAssis2017}. This effect of migration of charges, from one emitter to the others or to the substrate, is called ``mutual charge blunting" \cite{RFJAP2016} and it is also an active topic of research that deserves attention in order to achieve a better understanding of the physics behind small clusters of emitters.

\section{Conclusions}

The aim of the present work was to study how the emitter's degradation, as a consequence of changing morphology, may affect the field emission performance in small clusters of emitters in a LAFE. Our motivation was based on the well known fact that the LAFE may operate under sufficiently high temperatures, still on the field emission regime, and electrostatic fields. These conditions may lead to processes occurring during field emission that, as a consequence, irreversibly changes the local site FEF by a tradeoff between changes in the morphology and ``shielding". Besides that, the border effects and manufacturing limitations to control the geometry of a real LAFE, may also provide a different performance than the expected one. In order to explore these issues, we have used conformal mapping to analytically study the interplay between morphology and shielding effects in the apex-FEF, for a small cluster consisting of two identical emitters. We have considered two-dimensional models (ridge emitters), which allow an \textit{analytical} treatment. This approach is convenient for obtaining theoretical results that may explain the physics underlying this kind of phenomenon in real emitters. The emitters considered here were represented by conducting profiles limited by triangular protrusions on a line, which were assumed to be under an uniform electrostatic field.
We do not intend to explain the physics of the degradation processes, which is full of quantum and statistical challenges out of equilibrium, and is much beyond the classical electromagnetic theory used here. We simply study the FEF at electrostatic equilibrium for a given shape and distance between the emitters, considering that different values of the parameters used here may refer to different configurations of the LAFE, approximately at equilibrium after some possible degradation.

Our results feature a reduction of the local FEF and its variation at a region very close to the top of the emitters, in the limit of small distances between these apexes. Thus, we provide a theoretical explanation for the experimentally observed saturation in the FN plots, by showing that it may occur as a consequence of the tradeoff between morphological and shielding effects, as it does in our model. The shape of our FEF plots frequently resembles the ones obtained for the FEF in other works in the literature concerning the CPEE \cite{Jensen2015,JensenAPL2015,ForbesAssis2017,FT2017JPCM}. This is a promising phenomenon for technological applications in self-oscillations of multi-tip resonators. Nevertheless, it is shown that the geometry of the ridge emitters considered here, is not able to present CPEE. The similarity between our plots and the CPEE ones is also a consequence of the tradeoff between proximity and shape of the emitters. This represents an advance in understanding the limits and the conditions necessary for occurrence of the CPEE.

Perspectives for this work may include consideration of other models/geometries for field emitters in a LAFE. These models/geometries may also be amenable to explore the CPEE conditions theoretically. Additionally, one may reinforce the theoretical explanation obtained here for the saturation on FN plots or find alternative reasons for that.


\section*{Acknowledgement}

The authors acknowledge the financial support from CNPq and CAPES (Brazilian Agencies). TAdA also thanks Richard Forbes for illuminating discussions about the physics behind small clusters of emitters.

\end{document}